\documentclass{JHEP3}
\usepackage{epsfig}

\newcommand{\be}{\begin{equation}}
\newcommand{\ee}{\end{equation}}
\newcommand{\bea}{\begin{eqnarray}}
\newcommand{\eea}{\end{eqnarray}}
\newcommand{\IR}{\mathbb{R}} 

\def\IZ{\relax\ifmmode\hbox{Z\kern-.4em Z}\else{Z\kern-.4em Z}\fi}
\newcommand{\IS}{{\bf S}}

\newcommand{\non}{\nonumber \\}

\def\bw{\bar{w}}

\def\merger{\mbox{merger}}
\def\stwosq{{\bf S}^2 \times {\bf S}^2}
\def\tnt{\mbox{ Ton TNT }}

\preprint{{\tt hep-th/0207037}}

\title{Explosive black hole fission and fusion in large extra dimensions}

\author{Barak Kol
\\
 Institute for Advanced Study \\
 Einstein Drive\\
 Princeton NJ 08540,
 USA\\
\email{barak@sns.ias.edu} }

\abstract{Black holes are the densest form of energy, and in the
presence of compact dimensions black objects may take one of
several forms including the black-hole and the black-string, the
simplest relevant background being $\IR^{3+1} \times \IS^1$.
Recent understanding of the first order nature of the transition
indicate a powerful ``hysteresis'' curve, where black objects may
undergo fusion or fission during a tachyonic decay with Planck
power and duration of the order of the size of the compact
dimension $L$. Such explosions which scale with $L$ could be test
signatures for large extra dimensions in either astronomical
observations or accelerators.}

\begin{document}



\section{Introduction}

Black holes are by definition the densest form of energy. Black
holes which are small enough to be on the human scale ($\sim 1$m)
attract special fascination. For example a black hole of radius
$9mm$ has the following striking properties: it is long lived, has
the same mass as the earth, attracts matter 1m away with an
acceleration of $4\,  \cdot 10^{13}\, g$, has Hawking temperature
$0.02\, K^{0}$ and emission power of $9.2\,  \cdot 10^{-18}\, W$.
Any process involving such objects would involve energies and
forces which are extremely high on a human scale.

On the other hand, extra dimensions are theoretically motivated by
Kaluza-Klein theory and String Theory, and have recently
attracted much attention in phenomenology. Various extra
dimensions scenarios were suggested but for the purpose of this
note it will be enough to consider 4 extended dimensions together
with a fifth periodically identified coordinate $z, \, z \sim z+
L$, and it is enough to know that ``large extra dimensions'' much
larger than the Planck length cannot be ruled out experimentally
today, for $L \lesssim \hbar /1 \mbox{ TeV}$ in ordinary circle
compactification\footnote{Notation - while $c$ is set to $1$,
both $\hbar$ and $G_4$ (the 4d Newton's gravitational constant)
will be shown explicitly in order to distinguish classical from
quantum processes. The (4d) Planck length is defined by
 $l_{p4}^{~2} \sim \hbar \, G_4$, and the Planck mass is $m_{p4} \sim \hbar /l_{p4}$.}
 or even as high as $L \lesssim 1 \mu \mbox{m}$ for
theories where gauge fields are confined to a brane localized in
the extra dimension and only gravity propagates in the extra
dimension.

In the presence of a compact dimension massive solutions of
General Relativity may take one of several forms including the
black-hole and the black-string. Comparison of the entropies of
the two solutions reveals that for low masses (Schwarzschild
radius smaller than $L$) the black hole is preferred, while at
large masses the black string is stable, and the black hole phase
is expected to disappear following the merger of the ``north'' and
``south'' poles. This leads one to suspect that a phase transition
occurs. Indeed some phenomenological consequences of this
transition were discussed in
\cite{Casadio:2000py,Casadio:2001dc,Casadio:2001zz,Casadio:2001wh,Casadio:2002yj}.
The purpose of this note is to stress the phenomenological
consequences of a recent qualitative understanding of the
transition \cite{Gubser,BKtopology_change} (based on
\cite{GL1,GL2,SusskindGW,HorowitzMaeda}) and give estimates of the
energy and power emitted during transition. See also
\cite{Argyres:1998qn,Emparan:2000mb,Emparan:2000rs,Dimopoulos:2001hw,
Giddings:2001bu,Giddings:2001ih,Eardley:2002re,Chamblin:2002ad}
for a non-representative list of recent papers on black holes and
and large extra dimensions/ brane worlds/ accelerator prospects.

This note is organized as follows: we start with a review of the
suggested phase diagram of \cite{BKtopology_change} in section
\ref{phase_section}, then the fusion and fission processes are
described in section \ref{fission_fusion} together with explicit
numerical estimates and we conclude with a applications and open
numerical questions in section \ref{discussion}.

\section{Phase diagram}
\label{phase_section}

Let us set-up the problem.\footnote{Taken from
\cite{BKtopology_change}.} Consider a static (no angular
momentum) black object in an $\IR^{3+1} \times \IS^1$
gravitational background, namely extended $3+1$ space-time with a
periodic fifth dimension which will be denoted by the $z$
coordinate. The system is characterized by 3 dimensionful
constants: $L$ the size of the extra dimension $z \sim z + L$ ,
$M$ the (4d) mass of the system measured at infinity of $\IR^3$,
and $G_5$ the 5d Newton constant. From these a single
dimensionless parameter can be constructed \be \label{mu_M}
 \mu=(G_5 M)/L^2, \ee
while the 4d effective Newton constant is given by $G_4=G_5 /L$.

The isometries of these solutions are $SO(3) \times U(1)$, where
the $SO(3)$ comes from the spherical symmetry in $\IR^3$ and the
$U(1)$ comes from time independence.\footnote{In the Lorentzian
solutions it is the non-compact version of $U(1)$.} The most
general metric with these isometries is \be \label{general metric}
 ds^2 = -e^{2\, A} dt^2 + ds^2_{(r,z)} + e^{2\, C} d\Omega^2, \ee
which is a general metric in the $(r,z)$ plane together with two
functions on the plane $A=A(r,z),\, C=C(r,z)$. The horizon is a
line determined by $e^{2\, A}=0$ and as usual $d\Omega^2=d
\theta^2 + \sin^2{\theta}\, d \phi^2$. In 4d one can choose a
coordinate system that is especially adapted to cylindrical
symmetry \cite{Weyl1917}, but this works best only in 4d.

At least two phases of solutions can be distinguished - a black
string with an $\IS^2 \times \IS^1$ horizon topology and a 5d
black hole with an $\IS^3$ horizon. The uniform black string
(figure \ref{setup1}) is given by the 4d Schwarzschild metric
with $z$ added as a spectator coordinate \be
 ds^2 = -(1-r_4/r)\, dt^2 + (1-r_4/r)^{-1}\, dr^2 + dz^2 +
 r^2\, d\Omega^2 \ee
 where the Schwarzschild radius is $r_4=2\, G_4\, M$.
\EPSFIGURE{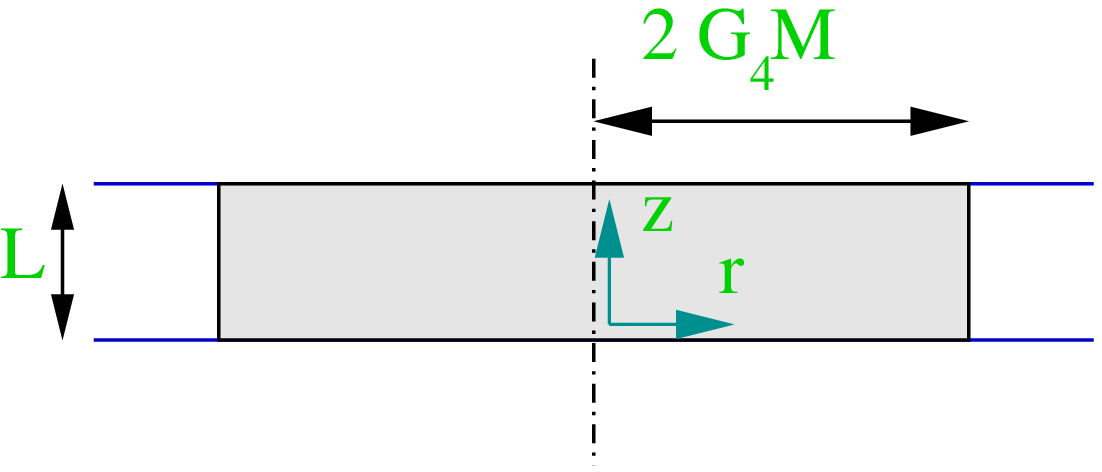}{The uniform black-string. $r$ is the radial
coordinate in $\IR^3$. \label{setup1}}

The small 5d black hole (see figure \ref{setup2})($r_5=\left(
(8/3\pi)\, G_5\, M \right)^{1/2} \ll L$) can be approximated by a
combination of two solutions. Denoting the distance from the
black hole by $\rho$, for $\rho \ll r_5$ it can be approximated
by the 5d Schwarzschild BH \be
 ds^2 = -(1-r_5^{~2}/\rho^2)\, dt^2 + (1-r_5^{~2}/\rho^2)^{-1}\, d \rho^2
 + \rho^2\, d\Omega^2_{~\IS^3} \ee
where $d\Omega^2_{~\IS^3}=d \chi^2 + \sin^2{\chi}\, d\Omega^2$.
For $\rho \gg r_5$ the Newtonian approximation is valid, and the
potential is proportional to the Green function \be
 V(w)={1 \over 4\, r} [\coth (w/2)+ \coth(\bw/2)] \label{5dNewtonian_potential} .\ee
 where $w:=2\, \pi \, (r+iz)/L$.
Indeed close to the source point $w \to 0$, the potential has the
expected 5d behavior $ V \sim (1/4r)[(2/w)+(2/\bw)] = L /(2\, \pi
(r^2+z^2))$, while for $\mbox{Re}(w) \to \infty$, the 4d behavior
is restored $\coth(w)=1+2 \exp (-2w), ~ V \sim (1/2r) +
O(\exp(-r)$.
 Presumably a full solution can be built perturbatively in
$r_5/L$ from these two approximations. Note that following the
equipotential surfaces of (\ref{5dNewtonian_potential}) one
encounters already a primitive version of a topology change -
close to the source the surface has an $\IS^3$ topology, then
there is one singular surface with a conic singularity (a cone
over $\IS^2$) after which the topology of the surfaces changes to
$\IS^2 \times \IS^1$.
 \EPSFIGURE{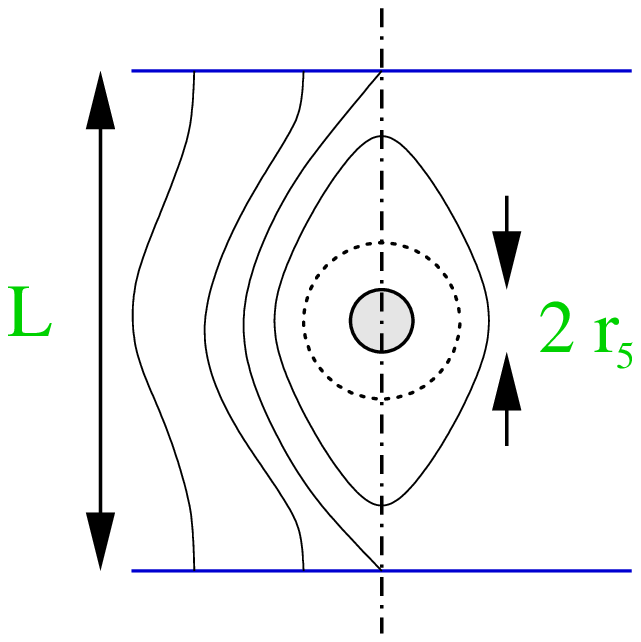}{The 5d black-hole
with Newtonian equipotential surfaces. \label{setup2}}

Comparing the areas of the two solutions one finds that for the
string the area is \be
 A_{string} = 4\, \pi\, r_4^{~2}\, L \sim \mu^2, \label{A_string} \ee
 while for the small BH \be
 A_{BH} \simeq 2\, \pi^2\, r_5^3 \simeq \mu^{3/2}. \label{A_BH} \ee
Hence, for small $\mu$ the black-hole is preferred, while for
larger $\mu$ it is not clear if the black-hole phase exists, but
even if it does then if its area would grow as in (\ref{A_BH}) it
would be dominated by the black-string. Therefore a phase
transition is expected from the outset.

We now turn to describe the phase diagram suggested in
\cite{BKtopology_change} -- figure \ref{phasediag} (some
refinements are in order near the merger point and will be
discussed below). The vertical axis is the parameter $\mu$, while
the horizontal axis is an order parameter\footnote{Even though
normally a phase diagram does not include order parameters.} of
non-uniformity $\lambda$. For small $\lambda$ it is defined by
$\lambda:=(1-r_{min}/ \bar{r})$ where $r_{min},\, \bar{r}$ are
the minimum and average respectively of $r=r(z)$, so that
$\lambda=0$ exactly when the solution is a uniform string. For
large $\lambda$ no precise definition is offered, but consider it
to continue to be a measure of non-uniformity even in the
black-hole phase.

\EPSFIGURE{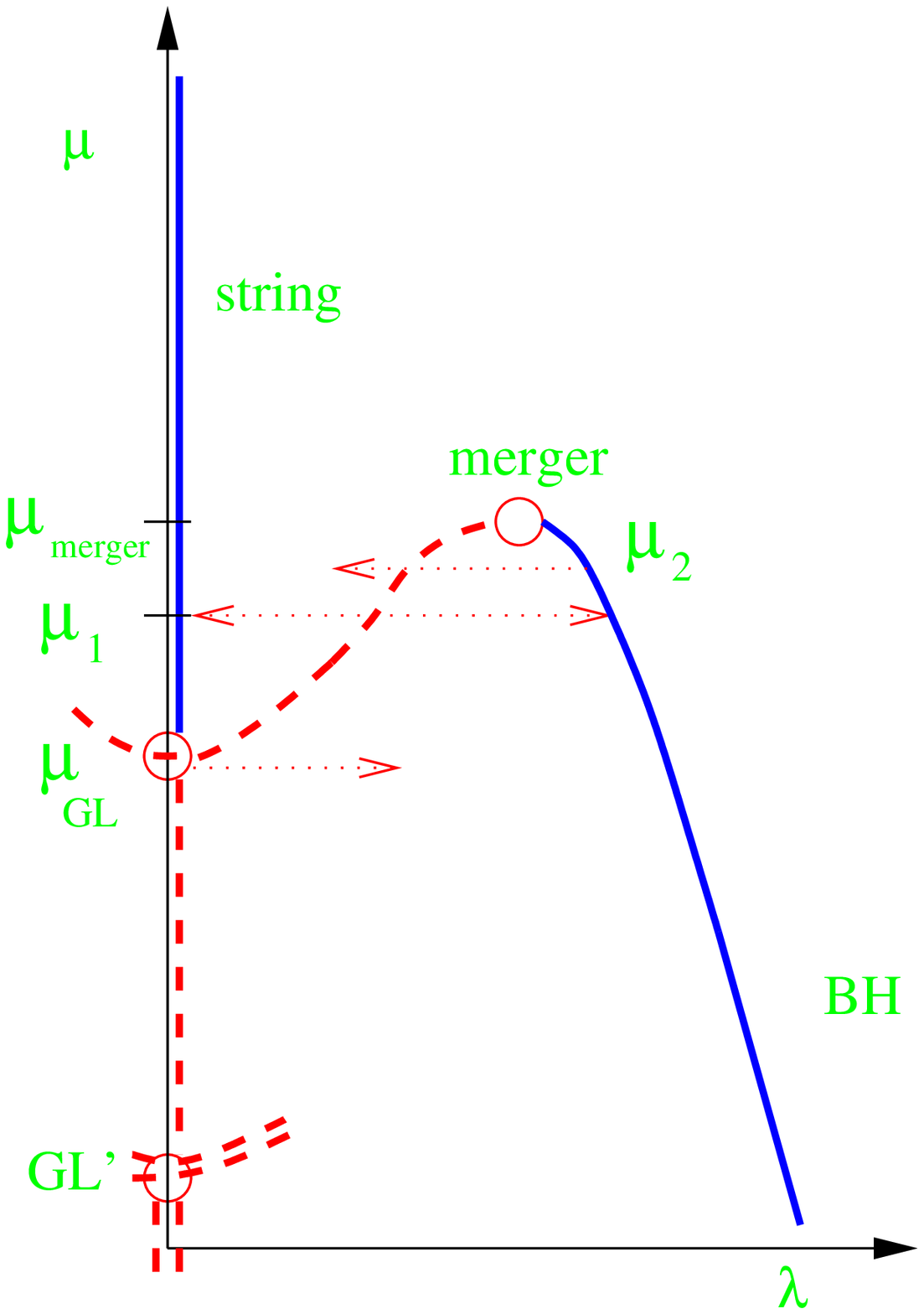,width=10cm}{A suggested phase diagram,
which will be refined further in the area of the merger
transition. $\mu$ is the dimensionless parameter, and $\lambda$
is an order parameter (measure of non-uniformity). Stable
(unstable) phases are denoted by solid (dashed) lines, while the
dotted lines denote transitions - a first order transition at
$\mu_1$ and a tachyonic decay from the two other points.
 \label{phasediag}}

The diagram contains two stable phases, the (uniform) black-string
and the black hole. The black string is stable for $\mu > \mu_1$
(the numerical value of $\mu_1$ is unknown) where a first order
transition occurs to an equal entropy black hole. Since the first
order phase transition requires tunneling, a big black string will
be long-lived and one may consider lowering $\mu$ further until
one gets to the Gregory-Laflamme point \cite{GL1,GL2} $\mu_{GL}
\simeq .070$ ( or $(L/r_4) \simeq 7.2$) where a certain metric
mode becomes marginally tachyonic signaling the breaking of the
translational symmetry along the $z$ axis. The end-point of this
decay is somewhat unclear and controversial -- while from the
current point of view and also according to \cite{GL1,GL2} the
end-point is the black-hole, it was also argued recently that the
end-point must be a stable non-uniform string
\cite{HorowitzMaeda}. At any rate, these distinctions do not
change the main result of this note, although they are important
for the details of the processes.

The unstable non-uniform string emanating from the GL point was
determined to be unstable by the analysis of Gubser \cite{Gubser}
and by some first results of a numeric study
\cite{Horowitz_talk,unpublished_simulation}. This determination
is crucial because it shows the transition to be first order and
explosive, rather than, say second order and mild. However, the
unstable phase never realizes physically.

The black-hole phase is stable for $\mu < \mu_1$ and long-lived
until the onset of a perturbative instability, which presumably
takes one to the uniform black-string phase. Now it is time to
explain the nature of the due refinement in figure
\ref{phasediag}. Actually the black hole must turn perturbatively
unstable already at $\mu_2 < \mu_{\merger}$. This is because the
cone over $\stwosq$ which is a central player in this transition
and locally models the merger and topology change is unstable
\footnote{Curiously this instability disappears for $d>10$
spacetime dimensions.}, and accordingly changes are expected in
the diagram in the vicinity of the merger point.

\section{Fission (evaporation) and fusion (accretion)}
\label{fission_fusion}

The terms ``fission'' and ``fusion'' refer to the following
process: we are interested in two kinds of transitions -
black-hole (BH) $\leftrightarrows$ black-string (see figure
\ref{transitions}). In ``black-string $\to$ BH'', the string
fissions into a BH, while in the ``BH $\to$ black-string'' the BH
fuses itself into a string.

\EPSFIGURE{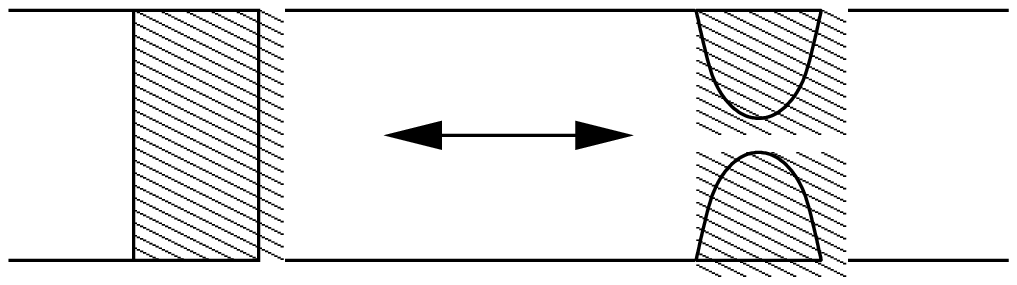,width=7cm}{The possible transitions
between black string and black hole (note that the vertical
direction is periodic). $\to$ is ``fission'' while $\leftarrow$ is
``fusion''. \label{transitions}}

The system was seen to include a typical first order transition.
In particular it traces a sort of ``hysteresis'' curve, with two
possible tachyonic decays. These decays are bound to release
energy in the form of gravitational waves, with charateristic
energy and time determined by their classical nature. Since at
transition the system has only one energy scale (classically) $M
\simeq L/G_4 \simeq L^2/G_5$ the released energy must be
proportional to it $\Delta M = \eta\, M$, with some efficiency
coefficient $\eta<1$ where $\eta$ may differ for the two
processes BH $\to$ black-string or black-string $\to$ BH, and it
can be expected to equal a few percent by analogy with black hole
collision simulations (see \cite{Lehner_review} and references
therein). For relevant $L$'s this can be restated as \bea
\label{energy}
 \Delta M &=& 1.21\,  \cdot 10^{38} J\; {\eta \over \mu}\, {L \over 1\, \mu m} =
  2.89\,  \cdot 10^{28} \tnt\; {\eta \over \mu}\,  {L \over 1\, \mu m} \non
          &=& 2.40\,  \cdot 10^{25} J\; {\eta \over \mu}\,  {L \over  \hbar\, 1\, \mbox{TeV}^{-1}} =
          5.71\,  \cdot 10^{15} \tnt\; {\eta \over \mu}\,  {L \over \hbar\, 1\, \mbox{TeV}^{-1}} \eea
where $1\tnt := 4.2$ GJoule, and for the black string
$\mu=\mu_{GL}=0.070$ while for the black hole $\mu \sim 1
>\mu_{GL}$.

The time scale for decay is again determined by the single
classical time scale present \be \label{time}
 \tau \sim L, \ee
 and so the power of emitted radiation is \be \label{power}
  P \sim \eta\, {1 \over G_4} \ee
It is interesting whether the emitted energy is originally
produced mostly in the form of gravitational waves, creating
other secondary particles, or whether other particles are
significantly produced.

Let us see how the fission of the black-string into a black-hole
fits with the Hawking evaporation process (see figure
\ref{evaporation}).
 \EPSFIGURE{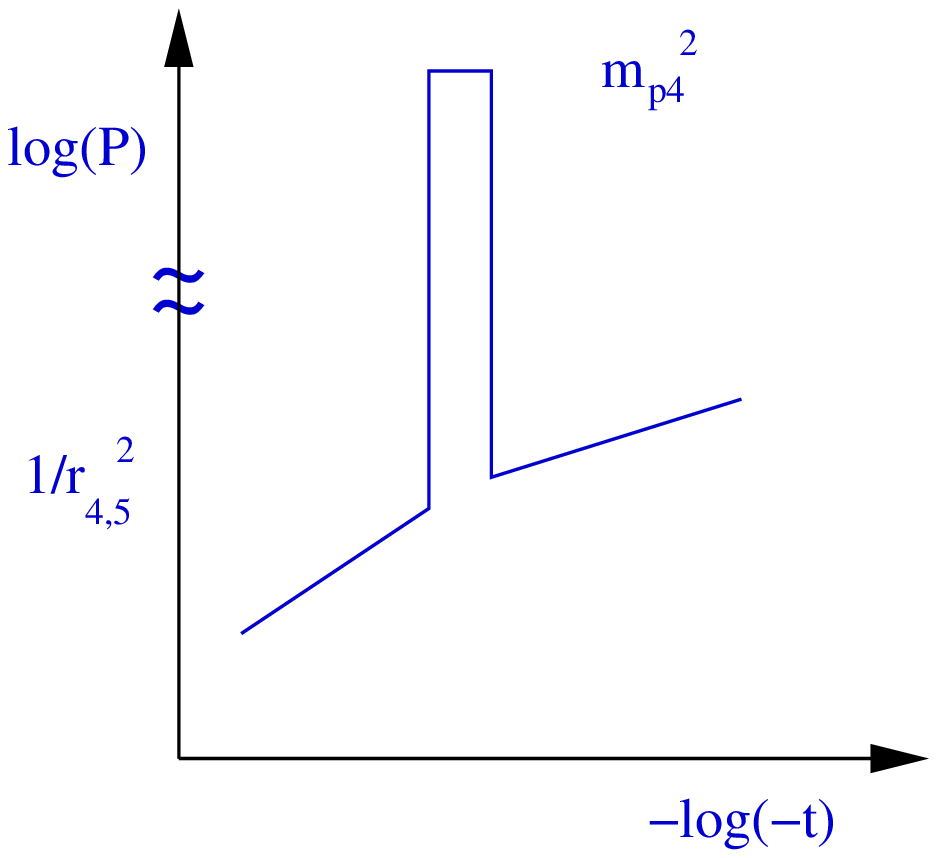,width=7cm}{A
typical evaporation process. The power is depicted as a function
of time on a log-log scale, and the kink between 4d and 5d
behaviour is over-accented. At transition an explosion occurs
with Planck power and duration of order $L$. \label{evaporation}}
  For $\mu > \mu_{GL}$ the emitted power is \bea
 P &=& \sigma\, A\, T^4 \simeq {\hbar \over (G_4\, M)^2} \simeq
 \non
 & \simeq & {\hbar \over r_4^{~2}}= 2.2 \cdot 10^{16}\, W\, \left({\hbar\, 1 \mbox{TeV}^{-1} \over
 r_4}\right)^2 \non
 & & \eea
where $\sigma=\pi^2\, k^4/(60\, \hbar^3\, c^2)=$\\ $5.67\,  \cdot
10^{-8} M\, m^{-2}\, K^{-4}$ is the Stefan-Boltzmann constant.
 Notice that this is a quantum process suppressed by a power of
$\hbar$. When $\mu_{GL}$ is reached the black string destabilizes
and collapses (``fission'') emitting at Planck powers
(\ref{power}). As a 5d black hole it continues to evaporate
though with a different rate $\sim \hbar/r_5^{2} \sim \hbar/
(G_5\, M)$. In the last stages of the evaporation for $M \sim
m_{p4}$ the power gets close to Planck power again but only over
Planck periods, so the total emitted energy is much smaller. Thus
the phase transition stands out as the most exothermic part of
the evaporation process.

Going in the opposite direction, a black hole may be ``charged''
up to $\mu_2$, say by an accretion disk, where it becomes
unstable and collapses to a string (``fusion''). The energy and
power parameters of this explosion are very similar to the
parameters for evaporation (\ref{energy},\ref{time},\ref{power}).
The ensuing evolution after the explosion will depend on the
balance of evaporation versus accretion.

\section{Discussion}
\label{discussion}

It would be interesting to apply these explosions (or their
absence) as a {\it test} or {\it signature for extra dimensions}.
As we saw, any Hawking evaporation process necessarily goes
through a fission explosion, the explosion being bigger the
bigger $L$ is. Fusion could also occur naturally for a growing
small black hole fed by an accretion disk. In relation to
astronomical observations one could estimate the chances for
small black holes to exist and explode within a time and space
window such that they would be noticable. One could also consider
implications for scenarios where black holes would be produced by
accelerators.

One cannot avoid considering {\it bomb construction} using either
fission or fusion ignoring the problems of creating and handling
small black holes. A black hole could be prepared close to its
critical size ready to fuse and then activated by adding matter
and reaching ``critical mass''. Timing the fission process is more
difficult, since the time for decay is determined by the Hawking
radiation which cannot be hurried artificially, but perhaps it is
possible to balance the Hawking radiation with incoming radiation
until activation. However, these bombs have a big disadvantage,
since if one is in possession of small black holes, one could
collide them with very similar explosions occurring both in
emitted energy and time scale (being a classical process), and
since igniting the phase transition seems to be the more
complicated process it would probably be disfavoured, and luckily
not used for destructive ends.

I would like to conclude by mentioning the basic dimensionless
constants of the problem which remain to be determined
numerically: $\eta_{1,2}$ the real-time efficiency for radiating
energy for either fission or fusion, $\mu_1$ the location of the
first order transition and $\mu_2$ the location of the onset of
the tachyon.

\vspace{0.5cm}

I would like to thank H. Davoudiasl for comments on a draft. Work
supported by DOE under grant no. DE-FG02-90ER40542, and by a
Raymond and Beverly Sackler Fellowship.


\end{document}